\documentclass[twocolumn]{autart}

\usepackage{natbib}
\usepackage{color}
\usepackage[dvips]{graphicx}
\usepackage{epsfig,amssymb,latexsym}
\usepackage{subfigure}
\usepackage{flushend}

\newtheorem{theorem}{Theorem}[section]
\newtheorem{lemma}{Lemma}[section]
\newtheorem{assumption}{Assumption}[section]
\newtheorem{remark}{Remark}[section]

\newcommand{\bea}{\begin{eqnarray}}
\newcommand{\eea}{\end{eqnarray}}
\newcommand{\be}{\begin{equation}}
\newcommand{\ee}{\end{equation}}
\newcommand{\ba}{\begin{array}}
\newcommand{\ea}{\end{array}}

\begin{document}

\begin{frontmatter}
\title{Direct Adaptive Controller for Uncertain MIMO Dynamic Systems with Time-varying Delay and Dead-zone Inputs \thanksref{footnoteinfo}}

\thanks[footnoteinfo]{This paper was not presented at any IFAC
meeting.
{\it E-mail addresses}: zjli@ieee.org (Zhijun Li), junfu@mit.edu (Jun Fu), cys@ustb.edu.cn (Changyin Sun).\\
This work was supported by the Natural Science Foundation
of China under Grants 61174045, 61473063 and 61125306, the Program for New Century Excellent Talents in University No. NCET-12-0195, Guangzhou Research Collaborative Innovation Projects (No. 2014Y2-00507), and National High-Tech Research and Development Program of China (863 Program) (Grant No. 2015AA042303).}

\author{Zhijun Li$^1$, ~~Ziting Chen$^1$, ~~Jun Fu$^{2,3}$, and ~~Changyin Sun$^4$}

\address{$^1$College of Automation Science and Engineering, South China University of Technology, Guangzhou, China, 510006.}
\address{$^2$Department of Mechanical Engineering, Massachusetts Institute of Technology (MIT), Cambridge, MA, USA, 02139.}
\address{$^3$State Key Laboratory of Synthetical Automation for Process Industries, Northeastern University, China, 110189.}
\address{$^4$School of Automation and Electrical Engineering, University of Science and Technology Beijing, China, 110083.}


\begin{abstract}
This paper presents an adaptive tracking control method for a class of nonlinearly parameterized MIMO
dynamic systems with time-varying delay and unknown nonlinear dead-zone inputs. A new high dimensional
integral Lyapunov-Krasovskii functional is introduced for the adaptive controller to guarantee global
stability of the considered systems and also ensure convergence of the tracking errors  to the origin.
The proposed method provides an alternative to existing methods used for MIMO time-delay systems with
dead-zone nonlinearities.
\end{abstract}

\end{frontmatter}

\section{Introduction}

In the past decades, control of multiple-input-multiple-output (MIMO) practical systems has attracted a
great deal of attention, such as flying robots (\cite{Ren:09}), biped robots (\cite{Ge:11}) and underwater
vehicle (\cite{Cui:10}). To characterize certain non-sensitivity for small control inputs of these MIMO
systems, dead-zone nonlinearities have to be considered otherwise it can severely limit system performances,
even leading to instability. To handle dead-zone nonlinearities, many control approaches have been developed
in the literature such as (\cite{Zhang:07}) and (\cite{Chen:13}). On the other hand, time-delay usually can
be encountered in these MIMO systems, dealt with by many authors, e.g., (\cite{Cui:10}) and references
therein. Time-delay, like the dead-zone nonlinearities, can also degrade the system performances or lead
to instability if ignored during the course of controller designs. Given the effects of the dead-zone
and time-delay, this paper considers a class of nonlinearly parameterized MIMO dynamic systems where
time-varying delay and unknown nonlinear dead-zone  are simultaneously taken into accounts.

To the considered problem in this paper, the most relevant papers are (\cite{Zhang:07}), (\cite{Chen:13}),
(\cite{Zhou:08}), (\cite{Shyu:2005}) and (\cite{Hua:08}). In (\cite{Zhang:07}), an adaptive neural control
was proposed for a class of uncertain MIMO nonlinear state time-varying delay systems with unknown nonlinear
dead-zones and gain signs, however, it only guarantees semiglobal stability of the closed-loop system.
Both (\cite{Chen:13}) and (\cite{Zhou:08}) used the backstepping techniques to construct the controllers.
However, the backstepping procedure is computationally time-consuming due to the computation of many
virtual controllers for the MIMO systems. The authors of (\cite{Shyu:2005}) proposed a decentralized
controller for large-scale systems with time-delay and dead-zone nonlinearities. However, in (\cite{Shyu:2005}),
the time-delay is constant and the parameters of the dead-zone are known. In (\cite{Hua:08}) the
authors constructed a novel Lyapunov function and then designed a smooth adaptive state feedback
controller for the time-delay system with dead-zone input. However, linear dead-zone was considered
and the tracking error can converge only within an adjustable region. Thus one may wonder if it is
possible to propose a new control method completely different from those existing ones to overcome
these disadvantages mentioned above? This paper provides an affirmative answer by introducing a new
high dimensional integral functional as a Lyapunov-Krasovskii function of the closed-loop systems.

From the motivation above and following our previous work (\cite{Ge:14}), this paper introduces a
new high dimensional integral Lyapunov-Krasovskii functional for a class of nonlinearly parameterized
MIMO systems with time-varying delay in states and unknown nonlinear dead-zones to achieve its tracking
control.

Compared to the existing results, the main contributions of this paper are: i) By proposing a Lyapunov-based adaptive control
structure, neither cancelation of the coupling matrix during linearizing the system nor conventional
backstepping techniques is needed; ii) By introducing a new high-dimensional integral Lyapunov function
in the control design, the process of controller design is simplified, i.e., it is unnecessary to
calculate the inverse of the unknown control gain matrix; iii) By the construction of the
Lyapunov-Krasovskii functional, the unknown time-varying delay in the upper bounding function of
the Lyapunov functional derivative can be easily eliminated; iv) The developed control strategy is applied to a 2-DOF robotic manipulator system and the comparative simulation studies demonstrate the superiority of the proposed method.

\section{Problem Statement and Assumptions} \label{s:sys}
Consider the following uncertain MIMO nonlinear time-delay system with dead-zone nonlinearities
\begin{eqnarray}\label{eq.input1.non1}
    \left\{\ba{ll}
    \dot {\mathbf{x}}_i=\mathbf{x}_{i+1},\\
    \dot {\mathbf{x}}_n=\mathbf{B}^{-1}(\mathbf{x})\left[\mathbf{F}(\mathbf{x}(t-\tau(t)))
                        +\mathbf{D}(t)+ \mathbf{u}\right],\label{eq.state}\\
    \mathbf{y}=\mathbf{x}_1, \ea
    \right. \label{eq.sys}
\end{eqnarray}
where $\mathbf{x}=[\mathbf{x}_1^T, \mathbf{x}_2^T, \dots, \mathbf{x}_n^T]^T\in \mathbb{R}^{nm}$ is the
state vector, $\mathbf{x}_i=[x_{i1}, x_{i2}, \dots, x_{im}]^T \in \mathbb{R}^{m}$ and $\mathbf{y}^{(i-1)}
=\mathbf{x}_i(i=1, 2,\dots, n)$; the nonlinear function $\mathbf{F}(\mathbf{x}(t-\tau(t))) \in \mathbb{R}^m$;
$\mathbf{B}(\mathbf{x}) \in \mathbb{R}^{m \times m}$ are unknown continuous bounded function matrix;
and the nonlinear function vector $\mathbf{D}(t)=[d_1,d_2. \dots, d_m]^T$ denotes the external disturbance.
$\mathbf{u}=[u_1,u_2, \dots, u_m]^T\in  \mathbb{R}^{m}$ is the output of the dead-zone control input and
satisfies that if $v_i\ge b_{ir}$, $u_i=  g_{ir}(v_i)$; if $b_{il} <v_{i}<b_{ir}$, $u_i= 0$; if $
v_i \le b_{il}$, $u_i=  g_{il}(v_i)$, where $v_i\in \mathbb{R}$ is the input to the $i$th dead-zone, and
$b_{il}$ and $b_{ir}$ are the unknown parameters of the $i$th dead-zone. In the paper, we consider the
dimension of $\mathbf{y}$ and $\mathbf{x}_i(i=1,2, \dots, n)$ are equal, therefore, $\mathbf{y}=\mathbf{x}_1$
holds.

The control objective is to find a control input $V =[v_1,v_2, \dots, v_m]^T$ such that the output of the
system tracks the desired trajectory $\mathbf{y}_d\in \mathbb{R}^m$ , while all the signals of the closed-loop
system are globally bounded.

\begin{remark}\label{ass.inertia}
The introduction of $\mathbf{B}(\mathbf{x})$ is for the physical meaning of mechanical system,
which can represent to the inertia matrix for the system.
\end{remark}

\begin{remark} \label{ass.positive}
The matrix $\mathbf{B}(\mathbf{x}) \in \mathbb{R}^{m \times m}$ is known to be either uniformly
symmetric positive definite or uniformly symmetric negative definite for all $\mathbf{x} \in \mathbb{R}^n$,
and have $m$ eigenvalues. Therefore, for the positive definite case, we have the following inequalities
\begin{eqnarray*} \lambda_{min}(\mathbf{B})\|x\|^2 \leq
x^T\mathbf{B}x \leq \lambda_{max}(\mathbf{B})\|x\|^2, \quad \forall
x \in \mathbb{R}^n,
\end {eqnarray*}
where $\lambda_{min}(\mathbf{B})$ and $\lambda_{max}(\mathbf{B})$ denote the minimum and maximum eigenvalues of
$\mathbf{B}$, respectively (\cite{Li:2008}), (\cite{Li:2013}).
\end{remark}

\begin{assumption}(\cite{Zhang:07})\label{ass.dead.out}
The dead-zone outputs $u_1, \ldots, u_m$ are not available and
the dead-zone parameters $b_{ir} \mbox{ and }b_{il}$ are unknown
constants, but their signs are known, i.e., $b_{ir}>0 \mbox{ and }
b_{il}<0, i=1,2,\ldots,m$.
The  growth of the $i$th dead-zone's left and right functions, $g_{il}(v_i)$
and $g_{ir}(v_i)$, are smooth, and there exist unknown positive constants
$k_{il0},k_{il1},k_{ir0}$ and $k_{ir1}$ such that
\begin{eqnarray} \label{eq.dead.dot.1}
0<k_{il0}&\le &g_{il}^{'}(v_i)\le k_{il1},\quad \forall v_i\in (-\infty,b_{il}],\\
\label{eq.dead.dot.2}
0<k_{ir0}&\le &g_{ir}^{'}(v_i) \le k_{ir1},\quad \forall v_i\in  [b_{ir},+\infty),
\end{eqnarray}
where $\beta_{i0}\leq  {\rm min}\{k_{il0}, k_{ir0}\} $ is a known positive constant, and
$g_{il}^{'}(v_i)={dg_{il}(z)}/{dz}|_{z=v_i}$ and $g_{ir}^{'}(v_i)={dg_{ir}(z)}/{dz}|_{z=v_i}$.
\end{assumption}

\begin{assumption} (\cite{Zhang:08})\label{ass.delay.bound}
The unknown state time-varying delays $\tau_i(t)$ satisfy $\dot{\tau}_i(t)\le \bar{\tau}_{max}<1$, $i=1, \cdots, m$,
with the known constants $\bar{\tau}_{max}$.
\end{assumption}

We know that there exist (\cite{Zhang:07}),
$\xi_{il}(v_i)\in (-\infty,b_{ir})$ and $\xi_{ir}(v_i)\in (b_{il},+\infty)$ such that
\begin{eqnarray} \label{eq.mean.value.l}
g_{il}(v_i)&=&g_{il}(v_i)-g_{il}(b_{il})= g_{il}^{'}(\xi_{il}(v_i))(v_i-b_{il}),\nonumber\\
&&\mbox{for}~~\xi_{il}(v_i) \in (v_i,b_{il}) \mbox{ \rm or } (b_{il},v_i),\label{eq.mean.value.2}\\
g_{ir}(v_i)&=&g_{ir}(v_i)-g_{ir}(b_{ir})=g_{ir}^{'}(\xi_{ir}(v_i))(v_i-b_{ir}),\nonumber\\
&&\mbox{for}~~\xi_{ir}(v_i) \in (v_i, b_{ir}) \mbox{ \rm or }(b_{ir},v_i).
\end{eqnarray}

Define vectors $ \Phi_i(t)=[\varphi_{ir}(t) ,\varphi_{il}(t) ]^T $ and $ K_i(t)=[g_{ir}^{'}(\xi_{ir}(v_i(t))),g_{il}^{'}(\xi_{il}(v_i(t)))]^{T}$
with
\begin{equation}
\varphi_{ir}(t)=\left\{
\begin{array}{lcl}
1 &\mbox{if}& v_i(t)>b_{il}, \\
0 &\mbox{if}& v_i(t)\le b_{il}, \\
\end{array}
\right. \nonumber \
\label{eq.char2.non}
\varphi_{il}(t)= \left\{
\begin{array}{lcl}
1 &\mbox{if}& v_i(t)<b_{ir},\\
0 &\mbox{if}& v_i(t)\ge b_{ir}.\\
\end{array} \right. \nonumber
\end{equation}

Based on Assumption \ref{ass.dead.out}, the dead-zone control input
can be rewritten as follows
\begin{equation}\label{eq.input.model.non}
u_i=K_i^T(t) \Phi_i(t) v_i+\xi_i (v_i),
\end{equation}
where
\begin{equation} \label{eq.distur.non}
\xi_i(v_i)=\left\{
\begin{array}{lcl}
- g_{ir}^{'}(\xi_{ir}(v_i))b_{ir}, &\mbox{if}&v_i\ge b_{ir},\\
-[g_{il}^{'}(\xi_{il}(v_i)) \\+g_{ir}^{'}(\xi_{ir}(v_i))]v_i, & \mbox{if}& b_{il} <v_i<b_{ir},\\
- g_{il}^{'}(\xi_{il}(v_i))b_{il}, &\mbox{if}& v_i\le b_{il},\\
\end{array} \right.
\end{equation}
and $|{\xi_i (v_i)}| \le p_i^\ast $, where $p_i^\ast $ is an unknown positive constant with
$p_i^*=(k_{ir1}+k_{kl1})\max\{b_{ir},-b_{il}\}$. Therefore, we have
\begin{eqnarray}
\mathbf{u}=K^T\Phi V+\Xi\label{eq.dead.output.u},
\end{eqnarray}
where $K={\rm diag}[K_i(t)]$, $\Phi={\rm diag}[\Phi_i(t)]$ and $\Xi =[\xi_1 (v_1), \dots,
\xi_m (v_m)]^T$.

Let $\mathbf{B}_d (\mathbf{x})\in \mathbb{R}^{m\times m}$ be a diagonal matrix with diagonal
elements $b_{dii}\neq 0(i = 1,2, \dots , m)$, then, there exists an unknown
matrix $\Delta_\mathbf{B}$ such that
$\mathbf{B}(\mathbf{x})=\mathbf{B}_d(\mathbf{x})+\Delta_\mathbf{B}$
is satisfied. We can rewrite (\ref{eq.state}) as
\begin{eqnarray}
    \mathbf{B}(\mathbf{x})\dot {\mathbf{x}}_n=\mathbf{F}(\mathbf{x}(t-\tau(t)))+
    \mathbf{D}(t)+\mathbf{u}.\label{eq.state2}
\end{eqnarray}
    Substituting (\ref{eq.dead.output.u}) and $\mathbf{B}(\mathbf{x})$ into (\ref{eq.state2}), we can obtain
\begin{eqnarray}
    \mathbf{B}_d(\mathbf{x})\dot {\mathbf{x}}_n
    =\mathcal{F}(\mathbf{x}(t-\tau(t)))+K^T\Phi V + \Delta_P,
    \label{eq.sys4}
\end{eqnarray}
where
\begin{eqnarray}
\mathcal{F}(\mathbf{x}(t-\tau(t)))&=&(I-\Delta_\mathbf{B}\mathbf{B}^{-1}(\mathbf{x}))
\mathbf{F}(\mathbf{x}(t-\tau(t)))\nonumber\\
    &=&[f_1, f_2, \dots,  f_m]^T
\end{eqnarray}
and
\begin{eqnarray}
\Delta_P=-\Delta_ \mathbf{B}
\mathbf{B}^{-1} K^T\Phi V+(I-\Delta_\mathbf{B}\mathbf{B}^{-1})
(\Xi+\mathbf{D})
\end{eqnarray}
are column vectors.

\begin{assumption}(\cite{Ge:99})\label{ass.approx}
Functions $b_{dii}$ and $d_i$ are continuous unknown. $b_{dii}$ and $d_i$
respectively satisfy $b_{dii}=W_{Bii}^T\Phi_{Bii}(\mathbf{x})$ and
$d_i \leq W_{di}^T \Phi_{di}(t)$ where $W_{Bii}\in \mathbb{R}^l$ is unknown bounded constant
parameter vectors, $\Phi_{Bii}(\mathbf{x}) \in \mathbb{R}^{l}$ is the known
continuous smooth bounded regressor vector, $W_{di} \in \mathbb{R}^l$ is a
vector of unknown bounded constant parameters, and $\Phi_{di}(t) \in \mathbb{R}^l$
is a vector of smooth bounded nonlinear function $\mathbf{x}$.
\end{assumption}

\begin{assumption}(\cite{Zhang:08})\label{ass.delay.func}
The unknown continuous functions $f_1 ,f_2, \ldots, f_m $ satisfy the inequality
\begin{equation}
|f_i|\le\sum_{k=1}^{m}\varrho_{ik}(\mathbf{x}_k(t-\tau_k(t)))
\end{equation}
with $\varrho_{ik}(x_k(t)) (i=1,2,\dots, m)$ being known positive continuous
functions.
\end{assumption}

\begin{lemma}(\cite{Ge:1998})\label{lem.transf-func}
Let $H(s)$ denote an $(n\times m)$-dimensional exponentially stable transfer function,
$r$ be the input and $ e=H(s)r$ be the output. Then $r\in L_2^m\bigcap L_{\infty}^m$ indicates that $e,
\dot{e}\in L_2^n\bigcap L_{\infty}^n$, $e$ is continuous, and $e\to 0$ as $t\to\infty$. Moreover,
if $r\to 0$ as $t\to\infty$, then $\dot{e}\to 0$.
\end{lemma}

\section{Control Design and Analysis}\label{sec.con.desi}

Define the filtered tracking error $s_i$ (\cite{Slotine:93})
\begin{eqnarray}
    s_i&=&e_i^{(n-1)}+\lambda_{i1}e_i^{(n-2)}+\dots+\lambda_{i,n-1}e_i,\label{eq.s}
\end{eqnarray}
where $e_i= y_i-y_{di}$, $e_i^{(n-1)}(i=1,2,\dots,m)$ is the $(n-1)$th derivative of
$e_i$, $\lambda_{i1},\dots,\lambda_{i,n-1}$ are positive constants and are appropriately
chosen coefficient vectors such that $e_i\rightarrow 0$ as $s_i\rightarrow 0$ (i.e.
$r^{m-1}+\lambda_{i1}r^{m-2}+\dots+\lambda_{i,n-1}$ is Hurwitz).

From (\ref{eq.s}), we have
\begin{eqnarray}
    \dot {\mathbf{S}}=\mathbf{B}_{d}^{-1}(\mathbf{x})\left[
    \mathcal{F}(\mathbf{x}(t-\tau(t)))+ K^T\Phi V+
    \Delta_\mathbf{P}\right]+\mathbf{\nu}, \label{eq.s3}
\end{eqnarray}
where $\mathbf{S}=[s_1,\dots,s_i,\dots,s_m]^T$ and $\mathbf{\nu}=[\nu_1,\dots,\nu_i,\dots,\\
\nu_m]^T$
with
\begin{equation}
\nu_i=-y_{di}^{(n)}+\lambda_{i1}e_i^{(n-1)}+\dots+\lambda_{i,n-1} .
\end{equation}

We now construct a new high dimensional Lyapunov-Krasovskii functional(see Eq. (\ref{lyapunov.V2})).
The first part of the Lyapunov-Krasovskii functional is chosen as
\begin{eqnarray}
    \mathbf{V}_1=\mathbf{S}^T\mathbf{B}_{\vartheta}\mathbf{S},\label{eq.v1}
\end{eqnarray}
where
\begin{eqnarray}\label{eq.Bv}
\mathbf{B}_{\vartheta}=\int_0^{1}\vartheta \mathbf{B}_{\alpha}d \vartheta
={\rm diag}\left[\int_0^{1}\vartheta \mathbf{B}_{\alpha ii}(\bar{\mathbf x}_i)d \vartheta\right]
\end{eqnarray}
with $\mathbf{B}_{\alpha}=\mathbf{B}_{d}\alpha={\rm diag}[b_{dii}\alpha_{ii}]_{m\times m}$ and
matrix $\alpha \in \mathbb{R}^{ m \times m}$.

 For easy analysis, we choose $\alpha_{11}=\dots=\alpha_{mm}$. By
exchanging $x_{ni}$ in $\mathbf x$ with $\vartheta s_i+\zeta_i(i=1,2,\dots, m)$, we define
$\bar{\mathbf x}_i=[\mathbf{x}_1^T,\mathbf{x}_2^T,\dots,\mathbf{x}_{n-1}^T,x_{n1},x_{n2},\dots,
\vartheta s_i+\zeta_i,\dots, x_{nm}]^T\in\mathbb{R}^{nm}$ where $\zeta_i=y_{di}^{(n-1)}-\xi_i$
with $\xi_i=\lambda_{i1}e_i^{(n-2)}+\dots+\lambda_{i,n-1}e_i$.
$\vartheta$ is a scalar and independent of $\bar{ \mathbf x}_i$.
We can choose suitable $\mathbf{B}_{d}(\mathbf{x})$ and $\alpha$, such that $b_{dii}\alpha_{ii}>0$.

\begin{remark}
We propose $\mathbf{B}_{\vartheta}$, which is a diagonal matrix, on the basis of weighted control Lyapunov
function (WCLF) defined in (Ge, Hang, Zhang (1999)). The diagonal element of $\mathbf{B}_{\vartheta}$ is
defined as $\int_0^{1}\vartheta \mathbf{B}_{\alpha ii}(\bar{\mathbf{x}}_i)d \vartheta$.
By introducing (19), we construct (18) as the first part of the
novel high dimensional Lyapunov-Krasovskii functional. Equation (\ref{eq.v1})
is positive definite in the filtered error
$\mathbf{S}$ and grows unbounded as $\|\mathbf{S}\|\rightarrow\infty$.
\end{remark}

Because $\mathbf{B}_{\vartheta}$ in (\ref{eq.v1}) depends on time $t$, the time derivative of $\mathbf{V}_1$ includes the differentiation of matrix $\mathbf{B}_{\vartheta}$ with
regard to time $t$. To facilitate computation of its derivative, according
to (\cite{Gentle:03}), we introduce a matrix operator for derivative operation of
matrix-value function with respect to time $t$, i.e., for a time-dependent matrix $A\in\mathbb{R}^{m\times n}$
and a vector $b(t)\in\mathbb{R}^l$, a matrix operator $\mathcal{M}_\partial(A,b)\in\mathbb{R}^{m\times n}$
is defined with the entry of its $i$th row and $j$th column being  $\mathcal{M}_{\partial ij}(A,b)
=\frac{\partial A_{ij}}{\partial b^T}\dot b$ with $i=1,2,\dots,m$ and $j=1,2,\dots,n$.

Differentiating
(\ref{eq.v1}) with respect to $t$ gives
\begin{eqnarray}
    \dot \mathbf{V}_1&=&2\mathbf{S}^T\mathbf{B}_{\vartheta}\dot\mathbf{S}+\mathbf{S}^T
    \mathcal{M}_\partial(\mathbf{B}_{\vartheta},\mathbf{S})\mathbf{S}\nonumber\\
    &&+\mathbf{S}^T\mathcal{M}_\partial(\mathbf{B}_{\vartheta},\mathbf{x})\mathbf{S}
    +\mathbf{S}^T\mathcal{M}_\partial(\mathbf{B}_{\vartheta},\zeta)\mathbf{S},
    \label{dot.v111}
\end{eqnarray}
where $\zeta=[\zeta_1,\zeta_2,\dots,\zeta_m]^T$,
$\mathcal{M}_\partial(\mathbf{B}_{\vartheta},\mathbf{S})\in\mathbb{R}^{m\times m}$,
$\mathcal{M}_\partial(\mathbf{B}_{\vartheta},\mathbf{x})\in\mathbb{R}^{m\times m}$
and $\mathcal{M}_\partial(\mathbf{B}_{\vartheta},\zeta)\in\mathbb{R}^{m\times m}$ are given below
\begin{eqnarray}
    &&\mathcal{M}_\partial(\mathbf{B}_{\vartheta},\mathbf{S})=
    {\rm diag}\left[\int_0^1\vartheta\frac{\partial\mathbf{B}_{\alpha ii}}{\partial
    s_i}\dot{s}_i d\vartheta\right],\label{eq.partial.S} \label{eq.partial.1}\\
    &&\mathcal{M}_\partial(\mathbf{B}_{\vartheta},\mathbf x)=
    {\rm diag}\left[\int_0^1\vartheta\sum_{j=1,j\ne i}^{nm}\frac{\partial\mathbf{B}_{\alpha ii}}{\partial
    x_j}\dot{x}_j d\vartheta\right],\label{eq.partial.bar.x}\\
    &&\mathcal{M}_\partial(\mathbf{B}_{\vartheta},\zeta)=
    {\rm diag}\left[\int_0^1\vartheta\frac{\partial\mathbf{B}_{\alpha ii}}{\partial
    \zeta_i}\dot\zeta_i d\vartheta\right],i=1,\dots,m.\label{eq.partial.v}
\end{eqnarray}

Let $\sigma=\vartheta s_i(i=1,2,\dots,m)$, we can obtain
\begin{eqnarray}
\frac{\partial\mathbf{B}_{\alpha ii}}
{\partial s_i}&=&\frac{\partial\mathbf{B}_{\alpha ii}}{\partial \sigma}\frac {\partial \sigma}{\partial s_i}
=\vartheta\frac{\partial\mathbf{B}_{\alpha ii}}{\partial \sigma},\\
\frac{\partial\mathbf{B}_{\alpha ii}}{\partial \vartheta}&=&\frac{\partial\mathbf{B}_{\alpha ii}}{\partial \sigma}\frac {\partial \sigma}{\partial \vartheta}
=\frac{\partial\mathbf{B}_{\alpha ii}}{\partial \sigma}s_i.
\end{eqnarray}

Noting that $\vartheta$ is a scalar and independent
of $\zeta_i$, and the fact $\dot \zeta_i=-\nu_i(i=1,2,\dots,m)$, we have
\begin{eqnarray}\label{eq.partial.n}
\int_0^1\vartheta\frac{\partial\mathbf{B}_{\alpha ii}}{\partial
\zeta_i}\dot\zeta_is_i d\vartheta=-\nu_i\int_0^1\vartheta\frac{\partial\mathbf{B}_{\alpha ii}}{\partial
s_i}d\vartheta.
\end{eqnarray}

Motivated by (\ref{eq.partial.1})-(\ref{eq.partial.n}), the following equations can be obtained
\begin{eqnarray}
    \mathbf{S}^T\mathcal{M}_\partial(\mathbf{B}_{\vartheta},\mathbf{S})\mathbf{S}
    &=&\mathbf{S}^T\left(\left.\left[\vartheta^2\mathbf{B}_{\alpha}\right]\right|_0^1-2
    \int_0^1\vartheta \mathbf{B}_{\alpha} d \vartheta\right)\dot{\mathbf{S}}\nonumber\\
    &=&\mathbf{S}^T\mathbf{B}_{\alpha}\dot\mathbf{S}-2\mathbf{S}^T\mathbf{B}_{\vartheta}
    \dot\mathbf{S},\label{eqution-parv}\\
    \mathbf{S}^T\mathcal{M}_\partial(\mathbf{B}_{\vartheta},\zeta)\mathbf{S}
    &=&\mathbf{S}^T\left(-\int_0^{1}
    \vartheta\frac{\partial\mathbf{B}_{\alpha}}{\partial \vartheta}d\vartheta\right)
    \mathbf{\nu}\nonumber\\
    &=&-\mathbf{S}^T\mathbf{B}_{\alpha}\mathbf{\nu} +\mathbf{S}^T\int_0^{1}\mathbf{B}_{\alpha}
    \mathbf{\nu}d\vartheta\label{eq.partinteg}.
\end{eqnarray}
By using the two equations above, we can rewrite (\ref{dot.v111}) as
\begin{eqnarray}\label{dot.V1}
    \dot \mathbf{V}_1&=&\mathbf{S}^T\mathbf{B}_{\alpha}\dot{\mathbf{S}}-\mathbf{S}^T
    \mathbf{B}_{\alpha}\mathbf{\nu}\nonumber\\
    &&+\mathbf{S}^T\left[\mathcal{M}_\partial(\mathbf{B}_{\vartheta},\mathbf x)\mathbf{S}
    +\int_0^{1}\mathbf{B}_{\alpha}\mathbf{\nu}d\vartheta\right].
\end{eqnarray}

Using (\ref{eq.s3}), we have
\begin{eqnarray}
    \dot\mathbf{V}_{1}&=&\mathbf{S}^T\mathbf{B}_{\alpha}\mathbf{B}_d^{-1}(\mathbf{x})
    \big[\mathcal{F}(\mathbf{x}(t-\tau(t)))+ K^T\Phi V+\Delta_\mathbf{P}\big]\nonumber\\
    &&+\mathbf{S}^T\left[\mathcal{M}_\partial(\mathbf{B}_{\vartheta},\mathbf x)\mathbf{S}
    +\int_0^{1}\mathbf{B}_{\alpha}
    \mathbf{\nu}d\vartheta\right]\label{eq.dotintegv}.
\end{eqnarray}

Since matrices $\mathbf{B}_d$, $\alpha$ and $\mathbf{B}_d\alpha$ are symmetric,we have
\begin{equation}
\mathbf{B}_{\alpha}\mathbf{B}_d^{-1}(\mathbf{x})=\mathbf{B}_d(\mathbf{x})
\alpha\mathbf{B}_d^{-1}(\mathbf{x})=\alpha.
\end{equation}

Then, equation (\ref{eq.dotintegv}) could be rewritten as
\begin{eqnarray}
    \dot \mathbf{V}_{1}&=&\mathbf{S}^T\alpha \left[\mathcal{F}(\mathbf{x}(t-\tau(t)))+
    K^T\Phi V + \Delta_\mathbf{P}\right] \nonumber\\
    &&+\mathbf{S}^T\left[\mathcal{M}_\partial(\mathbf{B}_{\vartheta},\mathbf x)\mathbf{S}
    +\int_0^{1}\mathbf{B}_{\alpha}
    \mathbf{\nu}d\vartheta\right]\label{eq.dotintegv2}.
\end{eqnarray}

From Assumption \ref{ass.approx}, we can rewrite (\ref{eq.dotintegv2}) as
\begin{eqnarray}
    \dot{\mathbf{V}}_1=\mathbf{S}^T \alpha [ \mathcal{F}(\mathbf{x}(t-\tau(t)))+
    W^T \Psi + K^T\Phi V +\Delta_\mathbf{P}], \label{eq.dotvf-1}
\end{eqnarray}
where $z =[\mathbf{x}^T, \bar{\mathbf{x}}^T, \mathbf{S}^T, \mathbf{v}^T]^T$, $W=
{\rm diag}[W_{Bii}]$,  $\Phi_B={\rm diag}[\Phi_{Bii}]$ and
\begin{equation}
\Psi(z) = \int_0^{1} \vartheta\mathcal{M}_\partial(\Phi_B,\mathbf x)\mathbf{S}d\vartheta
+\int_0^{1}\Phi_B\nu d\vartheta.
\end{equation}

Then, given Assumption \ref{ass.delay.func} and Lemma 2.1 in (\cite{Ge:14}),
it is easy to rewrite (\ref{eq.dotvf-1}) as follows
\begin{eqnarray}\label{eq.dotvf-2}
    \dot \mathbf{V}_1&=&\mathbf{S}^T\alpha\Big[ W^T \Psi(z)+K^T\Phi V+\Delta_\mathbf{P}
    \Big ]\nonumber\\ &&+\mathbf{S}^T \alpha  \mathcal{F}(\mathbf{x}(t-\tau(t))) \nonumber\\
    &\leq& \mathbf{S}^T \alpha \Big[W^T \Psi(z)+K^T\Phi V \Big]+\|
    \mathbf{S}^T \alpha \|( \gamma_1+\gamma_2 \|V\| )\nonumber \\
    && + \frac{m}{2}\|\mathbf{S}^T \alpha\|^2+ \frac{1}{2}
    \sum_{j=1}^m\sum_{k=1}^m  \varrho_{jk}^2(\mathbf{x}_k(t -\tau_k(t))).
\end{eqnarray}

\begin{remark}
 According to Remarks \ref{ass.inertia}  and \ref{ass.positive}, we can obtain that the
 bounded matrix $\mathbf{B}(\mathbf{x})$ is either uniformly symmetric positive definite or
 uniformly symmetric negative definite for all $\mathbf{x} \in \mathbb{R}^n$. Then the
 inverse $\mathbf{B}^{-1}(\mathbf{x})$ and unknown matrix $\Delta_{\mathbf{B}}$ should also
 be bounded. According to \emph{Assumptions \ref{ass.dead.out}} and \emph{Assumptions
 \ref{ass.approx}}, $\Xi$, $K$ and the disturbance vector $\mathbf{D}(t)$ are bounded.
 Therefore, the vectors $(I-\Delta_\mathbf{B}\mathbf{B}^{-1}(\mathbf{x}))\Xi$,
 $(I-\Delta_\mathbf{B}\mathbf{B}^{-1}(\mathbf{x}))\mathbf{D}(t)$ and matrix
 $\Delta_\mathbf{B}\mathbf{B}^{-1}(\mathbf{x})$ are bounded. As a result, there exist
 two positive parameters $\gamma_1$ and $\gamma_2$ such that
 \begin{eqnarray}
 &&\|(I-\Delta_{\mathbf{B}}\mathbf{B}^{-1}(\mathbf{x}))(\Xi+\mathbf{D}(t))\|\leq \gamma_1,\nonumber\\
 &&\|-\Delta_{\mathbf{B}}\mathbf{B}^{-1}(\mathbf{x})K \Phi V \|\leq \gamma_2\|V\|.\nonumber
 \end{eqnarray}
 Similar to literatures (\cite{Ge:14}) and (\cite{Xu:03}), $\gamma_1$ and $\gamma_2$ are
regarded as robust parameters in controller (\ref{eq.control}). In practice, we can constantly adjust the
two parameters until the proposed controller can stabilize the closed loop system.
\end{remark}

It is easy to check that $\Psi(z)$ is well-defined even if $\mathbf{S}$
approaches zero. We design an adaptive control
\begin{eqnarray}\label{eq.control}
    &&V=\mathbf{\Lambda}_{sgn}\mathbf{u}_1+\mathbf{u}_2,\\
    &&\mathbf{u}_1=-\beta_0^{-1}\left( (K_1+\frac{m}{2} \alpha)
    \mathbf{S}_{|\cdot|} +\hat{W}^T_{|\cdot|}\Psi_{|\cdot|}(z) +
    \Upsilon_{|\cdot|}\right),\nonumber\\\\
    &&\mathbf{u}_2=-\beta_0^{-1}\frac{\mathbf{S}}{\|\mathbf{S}\|}\rho,\label{control.u2}\\
    &&\rho=\left\{
    \begin{array}{lcl}
        \frac{\gamma_1+\gamma_2 \|\mathbf{u}_1\|}{1-\gamma_2
        \| \beta_0^{-1}\|},&\mbox{if}&\mathbf{S}\neq 0,\\
        0,&\mbox{if}&\mathbf{S} = 0,\\
    \end{array} \right.
\end{eqnarray}
where $\mathbf{\Lambda}_{sgn}={\rm diag}[sgn(s_i)]$; $(*)_{|\cdot|}$ denotes matrix or
vector that its every element is the absolute value of $(*)$'s corresponding element;
$\rho$ is positive when $\gamma_2 < {\| \beta_0^{-1}\|}^{-1}$; $\hat W$ is the estimate of $W$;
$K_1$ is a positive diagonal matrix; $\Upsilon$ will be defined later; and
$\beta_0={\rm diag}[\beta_{10}, \dots, \beta_{m0}]$ where $\beta_{i0}$ has been
defined in Assumption \ref{ass.dead.out}.

Since $K^T\Phi \geq  \beta_0>0$, when $\mathbf{S}\neq 0$, we can obtain
\begin{eqnarray}
    \mathbf{S}^T \alpha K^T\Phi V &\leq& - \mathbf{S}^T \alpha (K_1+\frac{m}{2} \alpha)
    \mathbf{S} - \mathbf{S}_{|\cdot|}^T \alpha(\hat{W}_{|\cdot|}^T\Psi_{|\cdot|}(z)\nonumber \\
    &&+\Upsilon_{|\cdot|} ) - \| \alpha \mathbf{S} \| \rho \nonumber \\
    &\leq& - \mathbf{S}^T \alpha (K_1+\frac{m}{2} \alpha) \mathbf{S} - \mathbf{S}^T
    \alpha ( \hat{W}^T\Psi(z) \nonumber \\
    &&+\Upsilon )-\| \alpha \mathbf{S} \| \rho,\label{eq.u2}
\end{eqnarray}
where we use the following facts
\begin{eqnarray}
&&-\mathbf{S}_{|\cdot|}^T \alpha\hat{W}_{|\cdot|}^T\Psi_{|\cdot|}(z)\le
-\mathbf{S}^T \alpha\hat{W}\Psi,\nonumber\\
&&-\mathbf{S}_{|\cdot|}^T \alpha\Upsilon_{|\cdot|}\le
-\mathbf{S}^T \alpha\Upsilon.\nonumber
\end{eqnarray}
Since $\|V\|\le\|\mathbf{u}_1\|+\|\mathbf{u}_2\|$, we have
$\|\mathbf{S}^T \alpha \|( \gamma_1+\gamma_2 \|V\|)-\| \alpha \mathbf{S} \| \rho\le0$.

Substituting (\ref{eq.u2}) into (\ref{eq.dotvf-2}) and noting that $\|V\|\le\|\mathbf{u}_1\|+\|\mathbf{u}_2\|$ give
\begin{eqnarray}
    \dot \mathbf{V}_1 &\leq& -\mathbf{S}^T \alpha K_1 \mathbf{S} - \mathbf{S}^T \alpha
    \tilde{W}^T \Psi(z)- \mathbf{S}^T \alpha \Upsilon \nonumber \\
    &&+ \frac{1}{2}\sum_{j=1}^m\sum_{k=1}^m  \varrho_{jk}^2(\mathbf{x}_k(t -\tau_k(t)))
    \label{eq.dotv11}.
\end{eqnarray}

Consider the following Lyapunov function candidate as
\begin{eqnarray}
    \mathbf{V}_2&=&\mathbf{V}_1+\mathbf{V}_a+ \sum_{j=1}^m \mathbf{V}_{U_j}(t),\label{lyapunov.V2}\\
    \mathbf{V}_a&=&\sum_{i=1}^m\frac{1}{2} \tilde{W}_i^T\Omega_i^{-1}\tilde{W}_i,
\end{eqnarray}
where $\tilde{W}_i=W_i-\hat{W}_i$.

The adaption law is designed as
\begin{equation}
\dot{\hat{W}_i} =\Omega_i\Psi_i(z)\alpha_{ii} s_i,
\end{equation}
where $\Omega_i>0(i=1,2,\dots,m)$ is a diagonal constant matrix to
be designed.

$\mathbf{V}_{U_j}(t)$ is introduced to overcome unknown time-delays
$\tau_1(t),\tau_2(t), \dots,\tau_m(t)$ and
defined as
\begin{eqnarray}
    \mathbf{V}_{U_j}(t) = \frac{1}{2(1-\bar{\tau}_{max})}\sum_{k=1}^m
    \int_{t-\tau_{k}(t)}^t\varrho_{jk}^2(\mathbf{x}_{k}(\tau))d\tau.
\end{eqnarray}
From the definition of $\mathbf{V}_a$, we have
\begin{equation}
\dot \mathbf{V}_a=\sum_{i=1}^m \tilde{W}_i^T\Omega_i^{-1}\dot{\tilde{W}_i}.
\end{equation}

The time derivative of $V_{U_j}(t)$ is
\begin{eqnarray}
\dot{\mathbf{V}}_{U_j}(t)&=&\frac{1}{2(1-\bar{\tau}_{max})}
    \sum_{k=1}^{m}\Big[\varrho_{jk}^2(\mathbf{x}_k(t))\nonumber \\
    &&-\varrho_{jk}^2(\mathbf{x}_k(t-\tau_k(t)))
    (1-\dot{\tau}_k(t))\Big].
\end{eqnarray}
Thus, the time derivative of $\mathbf{V}_2$ is
\begin{eqnarray}
\dot{\mathbf{V}}_2 &=& \dot \mathbf{V}_1+\dot \mathbf{V}_a+\sum_{j=1}^m
    \dot{\mathbf{V}}_{U_j}(t)\nonumber \\
    &\leq&\mathbf{S}^T \alpha(-K_1\mathbf{S}-\Upsilon) \nonumber \\
    &&+\frac{1}{2(1-\bar{\tau}_{max})}\sum_{j=1}^{m}\sum_{k=1}^{m}
    \varrho_{jk}^2(\mathbf{x}_k(t)),
\end{eqnarray}
where $ \Upsilon=[\Upsilon_1,\dots,\Upsilon_i,\dots,\Upsilon_m]^T$ with
\begin{equation}
\Upsilon_i=
\frac{1}{2(1-\bar{\tau}_{max})s_i\alpha_{ii}} \sum_{k=1}^{m}\varrho_{ik}^2(\mathbf{x}_k(t)).
\end{equation}

Noting that if $\Upsilon$ is utilized to construct the control law, controller singularity
may occur, since $({2(1-\bar{\tau}_{max})s_i\alpha_{ii}})^{-1}
\sum_{k=1}^{m}\varrho_{ik}^2(\mathbf{x}_k(t))$ is not well-defined at $s_i=0$. Therefore,
define $\Upsilon_i$ as follows:
\begin{equation}
\Upsilon_i=\left\{
\begin{array}{lcl}
    \frac{1}{2(1-\bar{\tau}_{max})s_i\alpha_i}\sum_{k=1}^{m}\varrho_{ik}^2(\mathbf{x}_k(t)),
    &\mbox{if}&s_i\neq0,\\
    0, &\mbox{if}&s_i=0.
\end{array}\right.
\end{equation}
Then, when $s_i \neq 0$,
\begin{equation}
\dot \mathbf{V}_2\leq  -\mathbf{S}^T\alpha  K_1\mathbf{S}\leq 0.
\end{equation}
To this end, our main result can be summarized as:

\begin{theorem}\label{the.stability.siso}
For the closed-loop system (\ref{eq.sys}) and (\ref{eq.control}),
under Assumptions \ref{ass.dead.out}-\ref{ass.delay.func}, for bounded initial
conditions, the tracking error $e_1$ converges to zero, and the overall closed-loop control system is globally stable in the
sense that all of the signals in the closed-loop system are globally bounded.
\end{theorem}

\begin{remark}
 From Lemma \ref{lem.transf-func}, it is obtained that the equation $\mathbf{S}=0$ defines a time-varying hyperplan in $\mathbb{R}^n$ on which the tracking error $e_1$ converges to zero asymptotically.
On the basis of this conclusion, we can obtain that $e_1$ converges to the origin if $\mathbf{S}$ converges
to zero asymptotically.
\end{remark}

\begin{remark}
In practice, to prevent chattering phenomena, $sgn$ function in (\ref{control.u2}) should be replaced by
${\rm sat}$ function, i.e.,
\begin{eqnarray}
\mathbf{sat}(\mathbf{S},\epsilon_s)=\left\{
\begin{array}{lcl}
    \mathbf{S}/\epsilon_s,
    &\mbox{if}& \|\mathbf{S}\| \leq \epsilon_s,\\
    \mathbf{S}/\|\mathbf{S}\|,
    &\mbox{if}& \|\mathbf{S}\| > \epsilon_s,\\
\end{array}\right.\nonumber
\end{eqnarray}
where $\epsilon_s$ is a small constant.
\end{remark}

\section{ Simulation Results}\label{sec.simu}

To validate the proposed method, we
consider the following 2-DOF robotic manipulator  system
\begin{eqnarray}
    \left\{\ba{ll}
    \dot {\mathbf{x}}_1=\mathbf{x}_{2},\\
    \dot {\mathbf{x}}_2={\mathbf{B}_d}^{-1}(\mathbf{x})\left[\mathcal{F}(
    \mathbf{x}(t-\tau(t)))+ \mathbf{D}(t)+K^T\Phi V\right],\\
    \mathbf{y}=\mathbf{x}_1, \ea
    \right.\nonumber
\end{eqnarray}
where ${\mathbf{B}_d}^{-1}(\mathbf{x})\mathcal{F}(\mathbf{x}(t-\tau(t)))
    =\mathbf{B}^{-1}(\mathbf{x})\mathbf{F}(\mathbf{x}(t-\tau(t)))$,
    $\mathbf{B}_d(\mathbf{x})={\rm diag} [b_{d11}, b_{d22}]$, $\mathbf{D}(t)=
[d_1, d_2]^T$,  $\mathbf{F}(\mathbf{x}(t-\tau(t)))=[a_{1}, a_{2}]^T$, $\mathbf{B}(\mathbf{x})=
[b_{11}, b_{12}; b_{21}, b_{22}]$, $\mathbf{x}_1=[q_1,q_2]^T$, $\mathbf{x}_2=[\dot q_1,\dot q_2]^T$,
$b_{d11}=2m_2l_1l_2(\cos q_2+2)$, $b_{d22}=m_2{l_2}^2(1+0.5\sin q_1)$,
$a_1=-m_2l_1l_2\dot q_1(t-\tau_2)\dot q_2(t-\tau_2)\sin q_2(t-\tau_1)
-m_2l_1l_2\dot q_2(t-\tau_2)(\dot q_1(t-\tau_2)+\dot q_2(t-\tau_2))
\sin q_2(t-\tau_1)$, $a_2=m_2l_1l_2{\dot q_1}^2(t-\tau_2)\sin q_2(t-\tau_1)$,
$b_{11}=(m_1+m_2){l_1}^2+m_2{l_2}^2+2m_2l_1l_2\cos q_2$,
$b_{12}=b_{21}=m_2{l_2}^2+m_2l_1l_2\cos q_2$, $b_{22}=m_2{l_2}^2$,
$d_1=(m_1+m_2)l_1g\cos q_2+m_2l_2g\cos(q_1+q_2)$ and
$d_2=m_2l_2\cos(q_1+q_2)$.

We choose robot parameters as $m_1=6$ kg,
$m_2=3$ kg, $l_1=0.8$ m, $l_2=0.4$ m
and $g= 9.81$ $m/s^2$ for numerical simulation. We consider the desired trajectory
$y_d=[\sin t, \cos t]^T$ and set the initial conditions $q(0) =
[0.3, 0.5]^T$ and $\dot q(0)=[0, 0]^T$. We choose $\hat{W}_1(0)=-0.1$ and $\hat{W}_2(0)=1$
as the initial value of adaption law. The design parameters of the above controller
are $\lambda_{11}=10$, $\lambda_{21}=5$, $\gamma_1=7.2$, $\gamma_2=0.2$,
$\Omega_1=0.02$, $\Omega_2=0.1$, $K_1={\rm diag}[10,5]$, $\beta_0={\rm diag}[0.3,1.1]$,
$\alpha={\rm diag}[14,14]$ and $\epsilon_s=0.2$. 
The parameters of the dead-zone are given as $g_{ir}(v_i)=k_{ir}(v_i-b_{ir})$
and $g_{il}(v_i)=k_{il}(v_i-b_{il})$ with the parameters of the dead-zones
$k_{1l} =0.5, k_{1r}=1.5, k_{2l} =1.5,k_{2r}=2.5, b_{1l}=-0.5,b_{1r} =0.5, b_{2l} =-2.5$ and $b_{2r} =2$.
The time-varying delays $\tau_1(t)=0.2(1.1+\sin t), \tau_2(t)=1-0.5\cos t$ and $\bar{\tau}_{max}=0.6$.

\begin{figure}
\centering
\includegraphics[width=90mm]{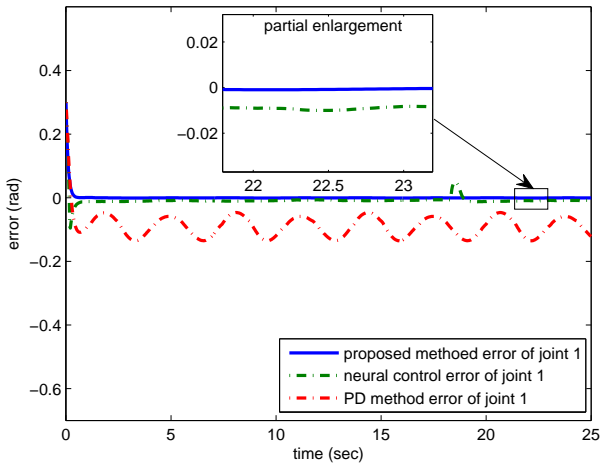}
\caption{Tracking error of joint 1.}\label{fig.err1}
\end{figure}

\begin{figure}
\centering
\includegraphics[width=90mm]{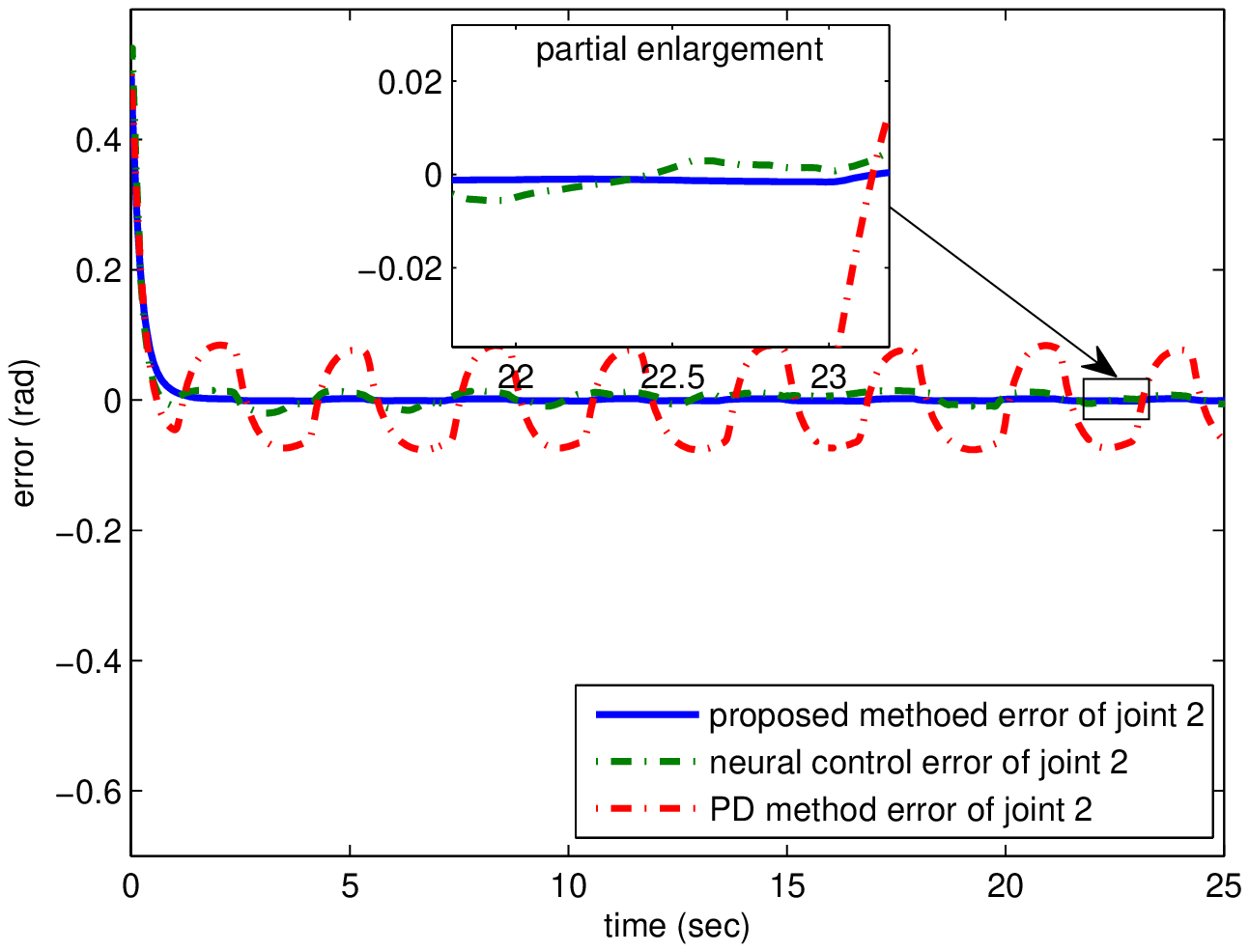}
\caption{Tracking error of joint 2.}\label{fig.err2}
\end{figure}

\begin{figure}
\centering
\includegraphics[width=90mm]{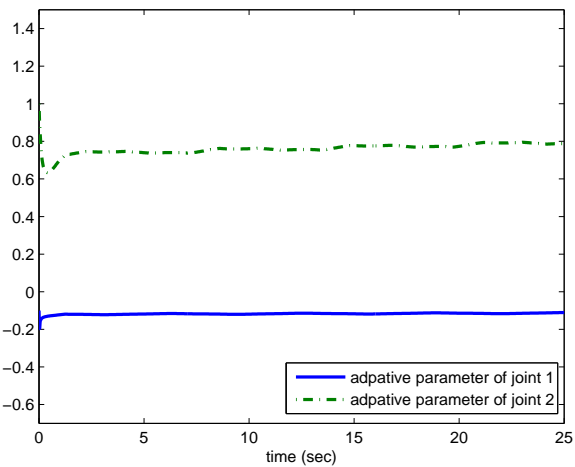}
\caption{Adaptive law $\hat{W}_i(i=1,2)$.}\label{fig.adaption}
\end{figure}

The tracking errors between the joint positions and their references are shown in Figs. \ref{fig.err1}--\ref{fig.err2}.
The time histories of the adaptive parameters are shown in Fig. \ref{fig.adaption}. These three figures
show good transient performances the proposed method achieves.

To show the advantages of the proposed method, we choose the conventional PD control
and the adaptive neural control proposed in (\cite{Zhang:07}) for comparisons under the same time-varying
delays and unknown dead-zones. As a traditional control method, the PD control can be written as
$V=-K_p e-K_d \dot e$. In the comparison simulation study, $K_p$ and $K_d$ are respectively set
to $K_p={\rm diag}[45,35]$ and $K_d={\rm diag}[10,10]$. For the adaptive neural control
proposed in (\cite{Zhang:07}),we also use two 3-layer neural networks
containing 10 hidden nodes to
approximate the unknown functions as done in (\cite{Zhang:07}). The controller parameters are chosen as
$\gamma_{10}=\gamma_{20}=3.5$, $c_{s_1}=c_{s_2}=0.1$, $\lambda_{11}=7.5$, $\lambda_{21}=5$,
$k_{11}=3$, $k_{21}=3.2$, $k_{13}=k_{23}=0.02$, $\eta_1=0.15$, $\eta_2=0.2$,
$\rho_1=\rho_2=2$, $\sigma_1=\sigma_2=0.01$, $\sigma_{w1}=\sigma_{w2}=0.1$,
$\sigma_{v1}=\sigma_{v2}=0.01$, $\Gamma_{w1}=\Gamma_{w2}={\rm diag}\{2.5\}$,
$\Gamma_{v1}=\Gamma_{v2}={\rm diag}\{15\}$ and $\bar \tau_{\max}=0.6$.

\begin{figure}
\centering
\includegraphics[width=90mm]{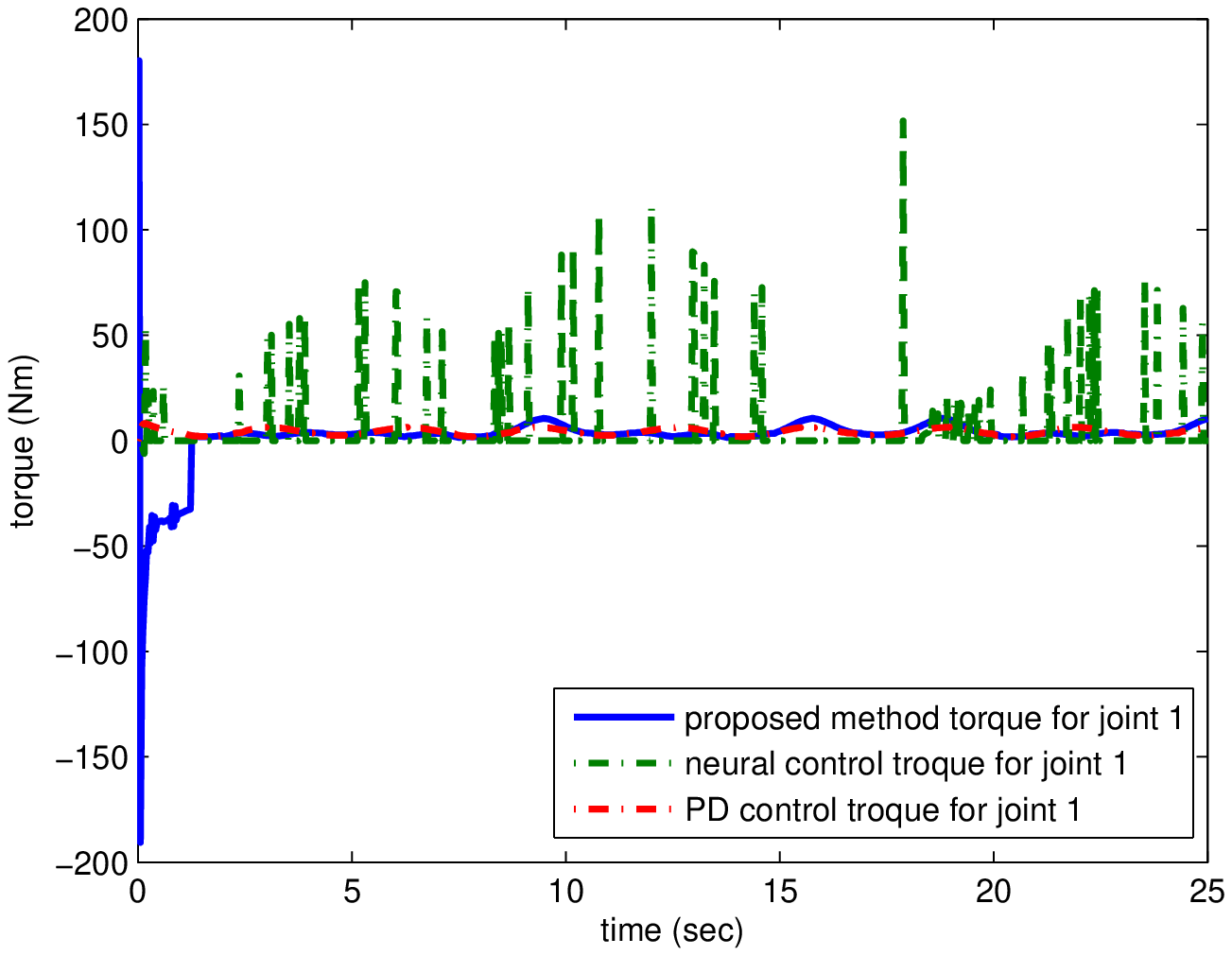}
\caption{Controller output $v_1$.}\label{fig.v1}
\end{figure}

\begin{figure}
\centering
\includegraphics[width=90mm]{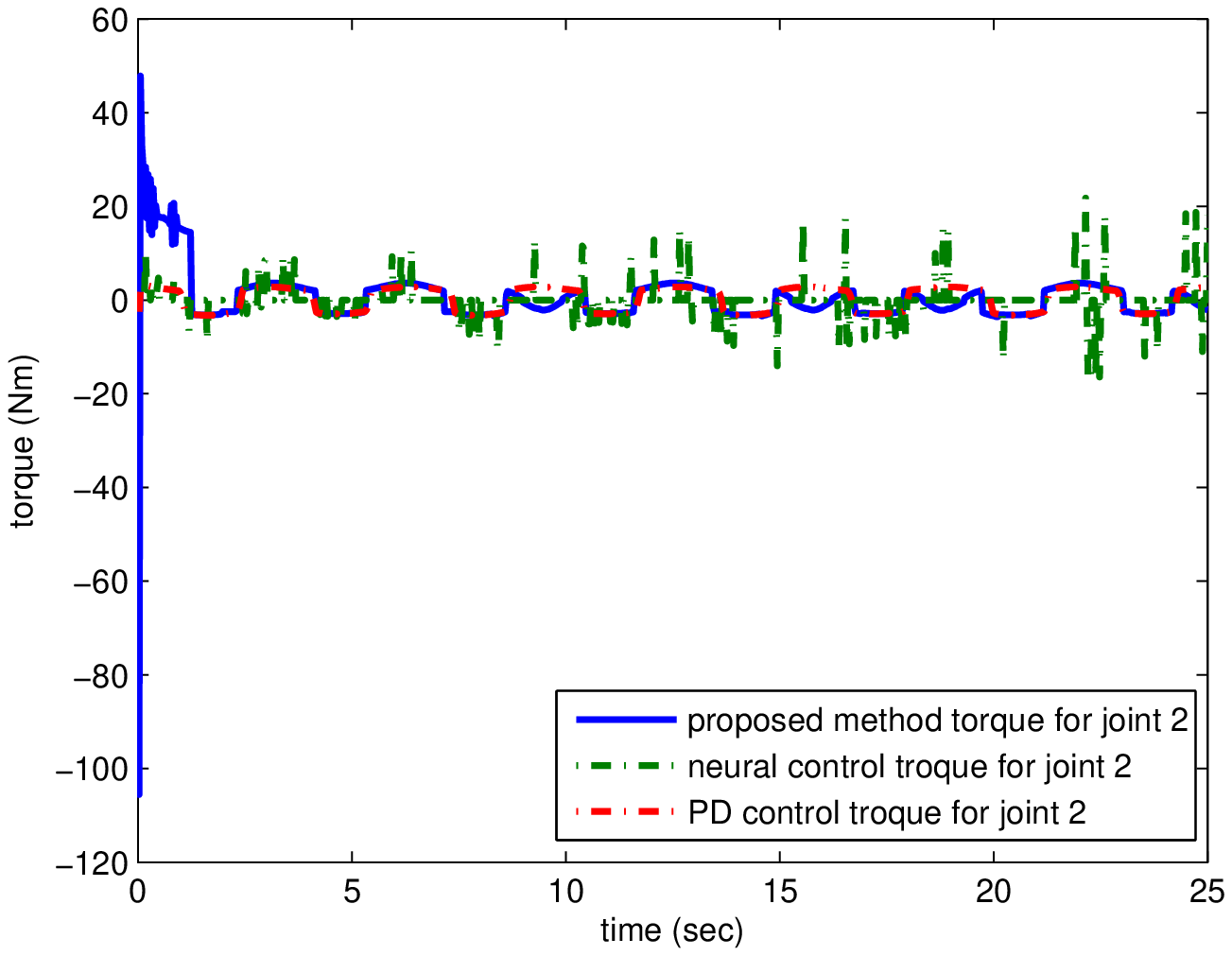}
\caption{Controller output $v_2$.}\label{fig.v2}
\end{figure}

The comparative simulation results are shown in Figs. \ref{fig.err1}--\ref{fig.err2}
and Figs. \ref{fig.v1}--\ref{fig.v2}. From Figs. \ref{fig.err1}--\ref{fig.err2}, we can see that the PD method cannot make the tracking
errors converge. From Figs. \ref{fig.v1}--\ref{fig.v2}, the control signals of the adaptive neural control proposed
in (\cite{Zhang:07}) can cause chattering phenomenon, which in practice can degrade system performances.
From the simulation results, our method can have better performances. Moreover, we construct the
controller mathematically by using the adaptive technique to deal with the uncertainty of the
considered system, instead of using neural networks approximation as in (\cite{Zhang:07}) and
(\cite{Zhang:08}). Furthermore, the method of (\cite{Zhang:08}) lies in the backstepping technique
which needs to construct $n$ controllers in the $n$ steps while our method only needs one step
in the sense of backstepping.

\begin{figure}
\centering
\includegraphics[width=90mm]{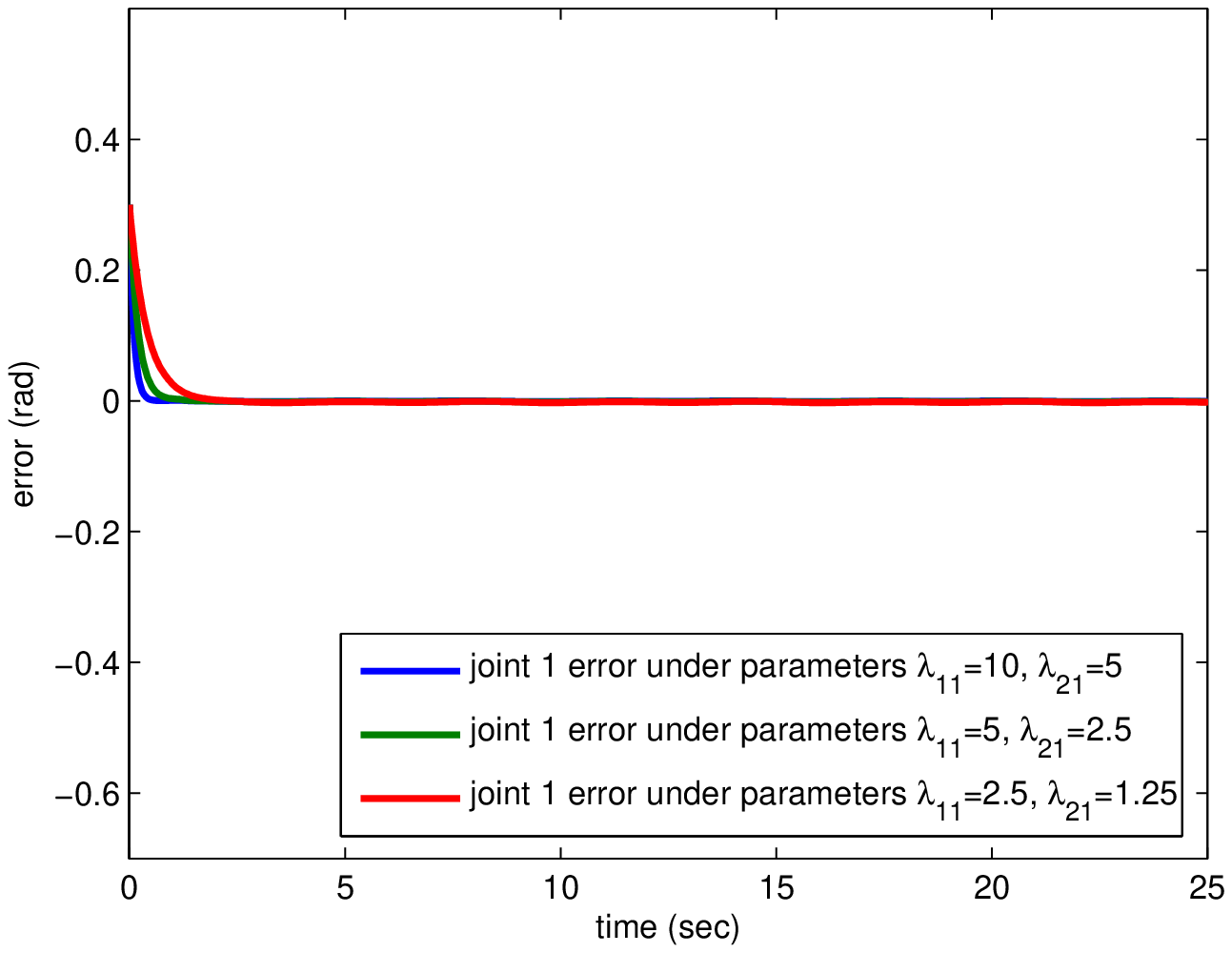}
\caption{Tracking error of joint 1 under different controller parameters.}\label{fig.err1.cmp}
\end{figure}

\begin{figure}
\centering
\includegraphics[width=90mm]{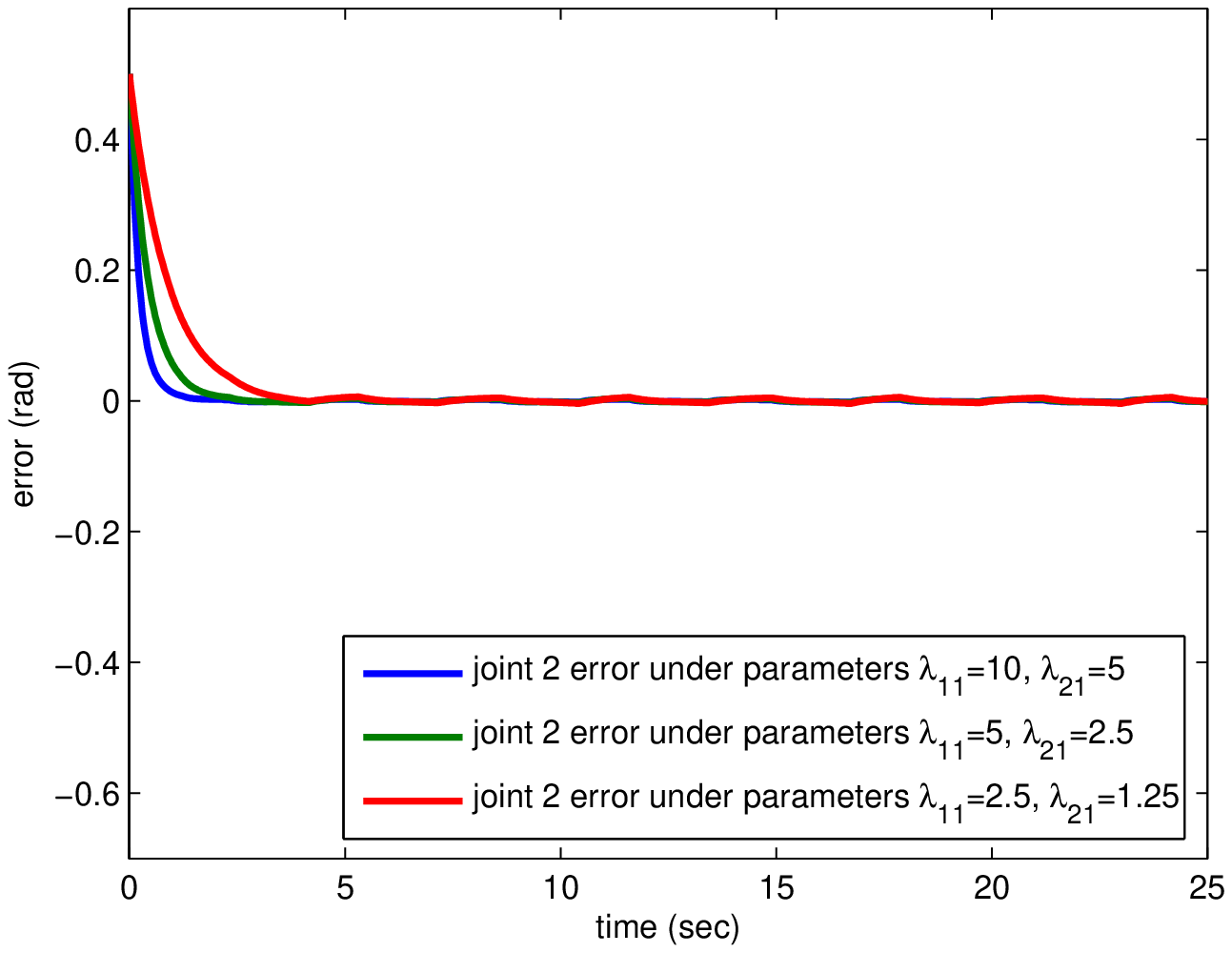}
\caption{Tracking error of joint 2 under different controller parameters.}\label{fig.err2.cmp}
\end{figure}

\begin{figure}
\centering
\includegraphics[width=90mm]{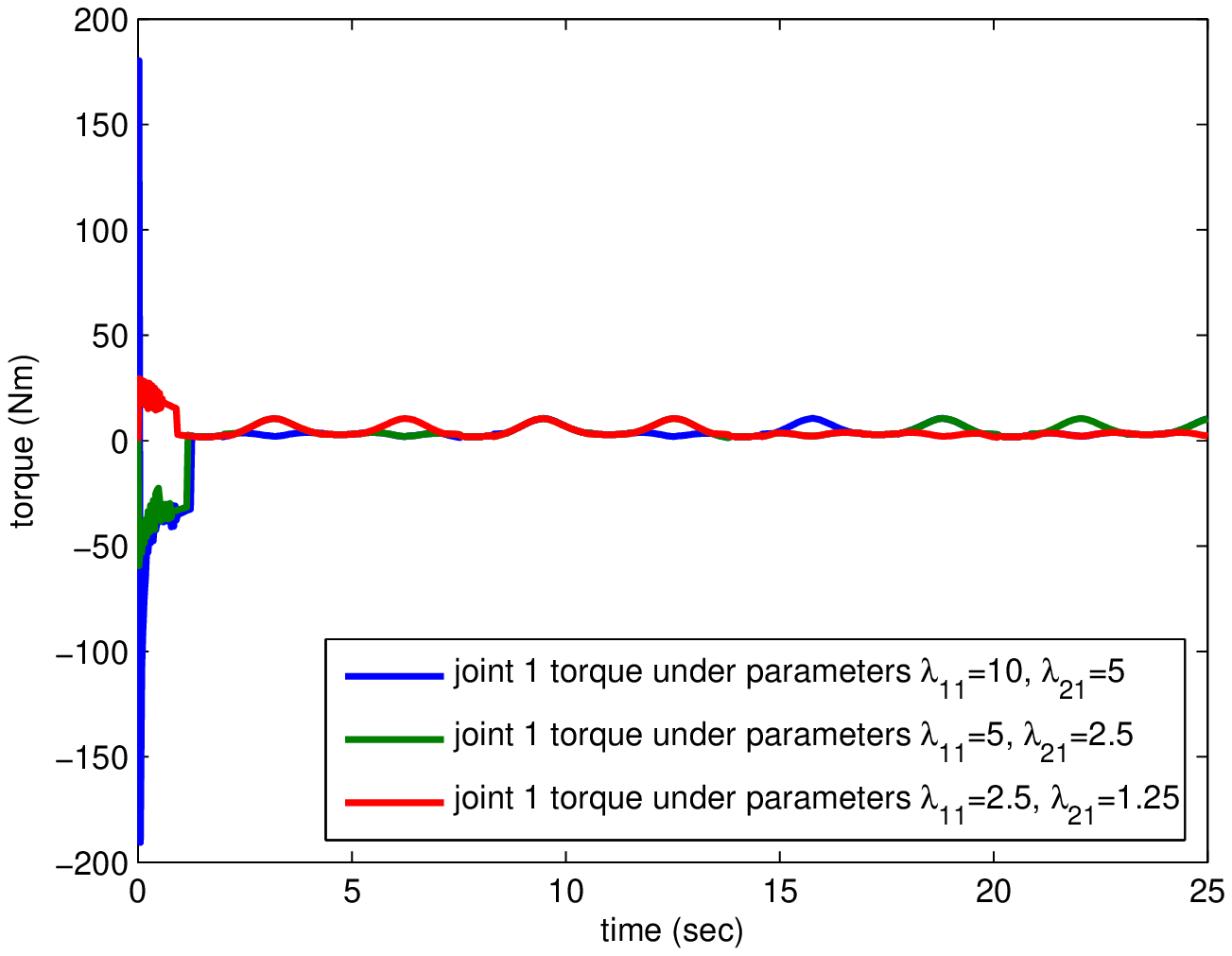}
\caption{Controller output $v_1$ under different controller parameters.}\label{fig.v1.cmp}
\end{figure}

\begin{figure}
\centering
\includegraphics[width=90mm]{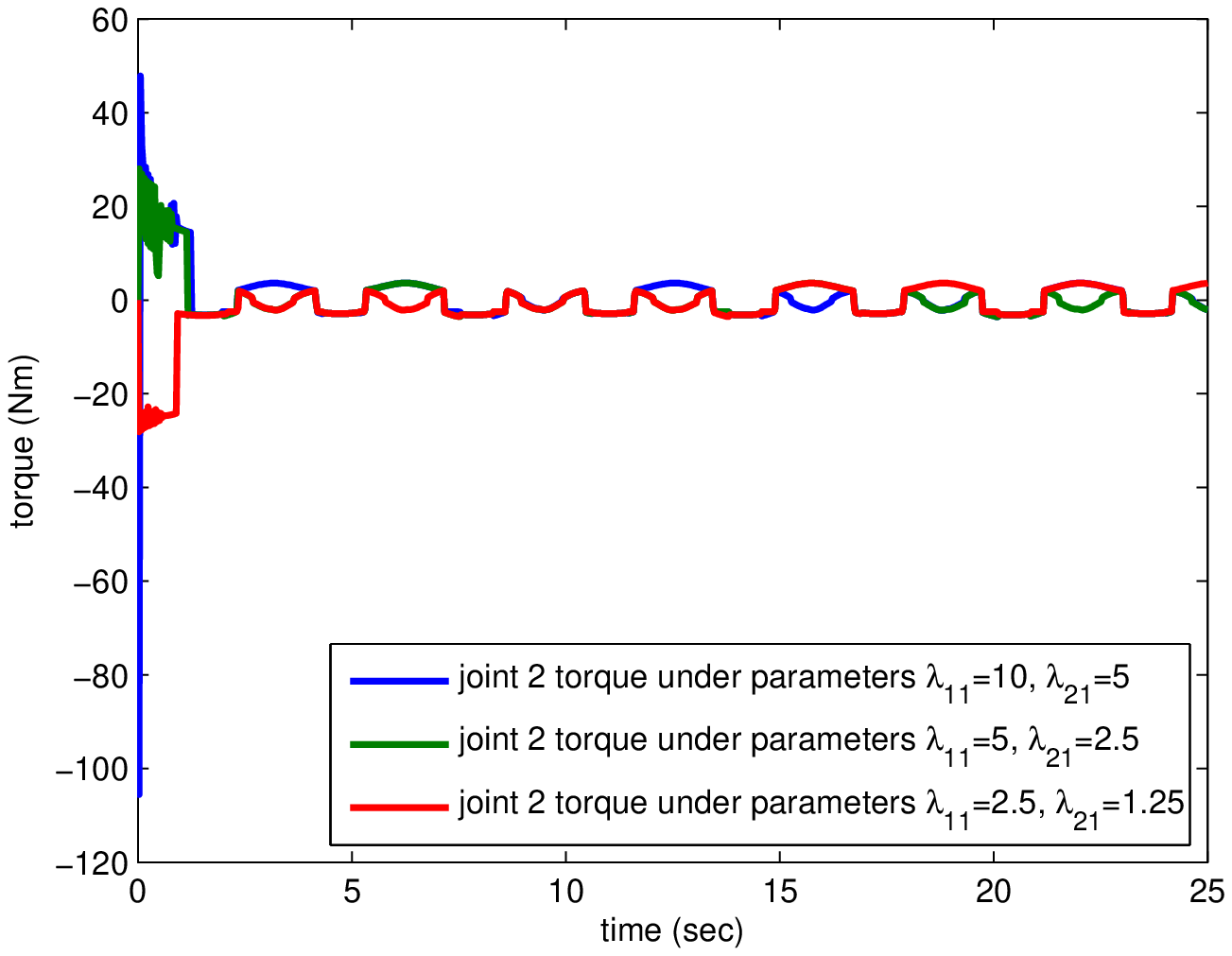}
\caption{Controller output $v_2$ under different controller parameters.}\label{fig.v2.cmp}
\end{figure}

In order to investigate the control performances of the proposed method under different
controller parameters, we also choose different parameters in the simulation.
Specifically, we select three pairs of different values of $\lambda_{11}$ and $\lambda_{21}$,
i.e., case 1: $\lambda_{11}=10$, $\lambda_{21}=5$, case 2: $\lambda_{11}=5$, $\lambda_{21}=2.5$,
case 3: $\lambda_{11}=2.5$, $\lambda_{21}=1.25$,  to observe how these two parameters affect
the control performances.  Figs. \ref{fig.err1.cmp}--\ref{fig.err2.cmp} and Figs.
\ref{fig.v1.cmp}--\ref{fig.v2.cmp}  show the tracking error trajectories
and controller output trajectories under different controller parameters, respectively. From these figures,
we can observe that the greater the values of $\lambda_{11}$ and $\lambda_{21}$ are, the faster
the convergence rate of tracking errors is, but accordingly the larger the control signals are
at the beginning of $t=0$.

\section{Conclusions}\label{sec.conclu}
In this paper, an adaptive control method is proposed for a class of uncertain MIMO nonlinear
time-varying delay systems with unknown nonlinear dead-zones. The design is based on the use
of a new high dimensional integral functional as a Lyapunov-Krasovskii function of the
closed-loop systems, with the advantages of global stability and convergence of the
tracking errors to origin.

\end{document}